\documentstyle[preprint,aps,epsf]{revtex}
\tightenlines
\newcommand{\lsim}{\mathrel{\lower4pt\hbox{$\sim$}}
\hskip-12.5pt\raise1.6pt\hbox{$<$}\;}
\newcommand{\gsim}{\mathrel{\lower4pt\hbox{$\sim$}}
\hskip-12.5pt\raise1.6pt\hbox{$>$}\;}
\begin{document}

\draft

\title{Improving constraints on $\tan\beta/m_H$ using
$B$$\to$$D\tau\overline{\nu}$}

\author{Ken Kiers\footnote{\tt{kiers@bnl.gov}\vspace{-.15in}} and 
Amarjit Soni\footnote{\tt{soni@bnl.gov}}}

\address{Department of Physics\\
Brookhaven National Laboratory, Upton, NY 11973-5000, USA}
\maketitle

\begin{abstract}

We study the $q^2$ dependence of the
exclusive decay mode $B$$\to$$D\tau\overline{\nu}$ in type II
two Higgs doublet models and show that this mode may be used
to put stringent bounds on $\tan\beta/m_H$.  There are currently
rather large theoretical uncertainties in the $q^2$ distribution,
but these may be significantly reduced by future measurements 
of the analogous
distribution for $B$$\to$$D(e,\mu)\overline{\nu}$.  We estimate
that this reduction in the theoretical uncertainties would
eventually (i.e., with sufficient data) allow
one to push the upper bound on $\tan\beta/m_H$
down to about $0.06$ GeV$^{-1}$.  This would represent
an improvement on the current bound by about a factor of $7$.
We then apply the method of optimized observables
which allows us to estimate the reach of an
experiment with a given number of events.  We thus
find that an experiment with, for example, $10^3$ events could set a 
$2\sigma$ upper bound on $\tan\beta/m_H$ of $0.07$ GeV$^{-1}$ or
could differentiate at the $4.6\sigma$ level
between a 2HDM with $\tan\beta/m_H = 0.1$ GeV$^{-1}$ and the SM.

\end{abstract}

\newpage

\section{Introduction}
\label{sec:1}

There has recently been considerable interest in constraining
the parameter space of type II two Higgs doublet models.
The main reason for this interest, of course, is that the 
Higgs sectors of minimal supersymmetric extensions of the standard
model are generically of this type~\cite{mssmhiggs}.  The charged Higgs
sectors of two Higgs doublet models (2HDM's) 
may be characterized by the ratio of the two Higgs' vacuum expectation values,
$\tan\beta$, and the mass of the charged Higgs, $m_H$.  
In this work we will investigate
how the exclusive decay channel $B\to D\tau\overline{\nu}$ may be used to 
place tight constraints on the ratio $\tan\beta/m_H$.  
This channel is expected to
have a branching ratio on the order of half a percent~\cite{wkn}, so that one
would expect on the order of $10^6$ such decays at the $B$ factories
which are currently under construction.

There already exist several constraints on $\tan\beta$ and $m_H$.
The most direct lower bound on the charged Higgs mass 
comes from the non-observation of charged Higgs pairs in $Z$ decays
and gives $m_H>44$ GeV~\cite{opaldelphi}.  
Another limit comes from top decays, which yield the bound
$m_H>147$ GeV for large $\tan\beta$~\cite{cdf}.
Finally, for pure type II 2HDM's one finds
$m_H>300$ GeV, coming from the virtual Higgs contributions to
$b\to s\gamma$~\cite{hewett}.  This
latter limit disappears in the context of supersymmetry since
the Higgs contributions to $b\to s \gamma$ can be cancelled by
other contributions~\cite{bsgammasusy}.
There are no experimental upper bounds on the mass of
the charged Higgs, but one generally expects to have
$m_H < 1$ TeV in order that perturbation theory remain
valid~\cite{maxmass}.
A lower limit may be placed on $\tan\beta$ 
by considering the branching ratio for $Z\to b\overline{b}$.
The resulting bound of $\tan\beta > 0.7$, obtained in
Ref.~\cite{grant}, coincides with the range generally
favoured by theorists in order that renormalization group
evolution drive electroweak symmetry breaking~\cite{dawson}.
For large $\tan\beta$ the most stringent constraints on $\tan\beta$
and $m_H$ are actually on their ratio, $\tan\beta/m_H$.  The current
limits come from the measured branching ratio for the 
inclusive decay $B\to X \tau\overline{\nu}$, giving
$\tan\beta/m_H < 0.46$ GeV$^{-1}$~\cite{aleph1}, 
and from the upper limit on the 
branching ratio for $B\to \tau\overline{\nu}$, giving
$\tan\beta/m_H < 0.38$ GeV$^{-1}$~\cite{l3}.
While both of these limits
are quoted as being at the $90\%$ confidence level, the latter
may be somewhat less constraining due to the uncertainties
in $V_{ub}$ and $f_B$.

Our main goal in this paper is to 
investigate the sensitivity of the exclusive
decay $B\to D\tau\overline{\nu}$ to the ratio $\tan\beta/m_H$.  
We will concentrate on the $q^2$ distribution for this decay
and discuss how the theoretical uncertainties in this distribution may
be minimized.  We also apply the optimized weighting 
procedure~\cite{atsoni,ggh} to this distribution in order
to derive quantitative estimates for the sensitivities of 
experiments with given numbers of events.  This procedure can be shown 
to give the smallest statistical uncertainty when analyzing the data
in a given experiment.  The present work complements the 
previous theoretical studies of the 
inclusive~\cite{kraw,kal,ghou1,hou,isidori,grosslig1,ghn,coarasa} and 
exclusive~\cite{ghou2,tanaka} semi-tauonic $B$ decays,
as well as those of the purely leptonic decays
$B\to\tau\overline{\nu}$~\cite{hou} and 
$B_c\to\tau\overline{\nu}$~\cite{du}.
These decays are attractive because the Higgs contribution
occurs at tree-level and cannot be cancelled by, for example, supersymmetric
loop effects.  Thus, the results of our analysis should
be applicable to any type II 2HDM~\cite{note:coarasa}  This situation 
may be contrasted with that in $b\to s\gamma$~\cite{bsgammasusy}.

A feature which is common to all of the tauonic and semi-tauonic $B$ 
decays is that the Higgs
contribution to the amplitude interferes destructively with that due
to the standard model (SM).  As a result, the corresponding 
integrated partial widths, plotted as functions of 
$\tan\beta/m_H$, tend to have minima around 
$\tan\beta/m_H \sim 0.2$ -- $0.3$ GeV$^{-1}$.  Most studies to date
have concentrated on the region to the right of the minimum, where
the Higgs contribution to the width begins to dominate over the SM 
contribution.  Indeed, the present experimental limits -- derived
using {\em integrated} partial widths -- correspond
to this region.  In order to use semi-tauonic decays
to probe values of $\tan\beta/m_H$
near and/or below the minimum, it will be extremely useful to have
detailed theoretical predictions for quantities beyond simply
the integrated partial widths.  This is because
the plots of the widths as
functions of $\tan\beta/m_H$ are generically relatively flat up to 
$\tan\beta/m_H \sim 0.4$ GeV$^{-1}$.
Several authors have suggested using the energy distribution
or longitudinal polarization of the $\tau$ in this regard,
but this may be difficult experimentally since two neutrinos
are always lost.  An alternative approach, which we will study in detail,
is to use the $q^2$ distribution.  A possible drawback of this
approach is that the $q^2$ distribution is very sensitive to theoretical
uncertainties in the shapes of the hadronic form factors.
As we shall see, however, the situation in the 
exclusive channel $B\to D\tau\overline{\nu}$
appears to be quite encouraging.  The reason for this
is that once the distribution for $B\to D (e,\mu)\overline{\nu}$ has
been measured, that for $B\to D\tau\overline{\nu}$ can be predicted
with relatively small theoretical uncertainties.
The resulting distribution is quite sensitive
to $\tan\beta/m_H$, even for relatively small values of this
ratio.  Furthermore, the $q^2$ distribution has a qualitatively different
shape for values of $\tan\beta/m_H$ above and below the ``critical value'',
$\tan\beta/m_H\sim 0.3$ GeV$^{-1}$.

We have chosen to focus on the decay
channel $B\to D\tau\overline{\nu}$ instead of on
$B\to D^*\tau\overline{\nu}$, even though the latter channel will
likely have a somewhat larger
branching ratio and may also be more accessible experimentally
(in analogy with the decays to the lighter leptons~\cite{neubert1})
than the former.
Our main motivation for considering $B\to D\tau\overline{\nu}$
rather than $B\to D^*\tau\overline{\nu}$ is simply that
the Higgs contribution has a much larger effect in the former case.
This feature has
already been noted in Ref.~\cite{tanaka} and is in part due to an
enhancement by a factor $(m_B+m_D)/(m_B-m_D)\sim 2$ in the effective
interaction.  As noted in Ref.~\cite{tanaka}, this enhancement effect
means that the exclusive $D$ channel is also more sensitive
than the {\em inclusive} channel, since the less-sensitive $D^*$ mode
tends to dilute the inclusive measurement.

The plan of the remainder of this paper is as follows.
We begin in Sec.~\ref{sec:2} by deriving the $q^2$ distribution
for $B\to D\tau\overline{\nu}$ in terms of the dimensionless
variable $t = q^2/m_B^2$.  In Sec.~\ref{sec:3} we estimate the theoretical
uncertainties in this distribution and in the 
integrated width once the distribution for 
$B\to D (e,\mu)\overline{\nu}$ has been measured.  Barring any
further input, these uncertainties would eventually limit the reach
of such an experiment.  In Sec.~\ref{sec:4} we apply the optimized
weighting procedure to the $q^2$ distribution and in
Sec.~\ref{sec:5} we present our conclusions.

\section{Calculation of the differential distribution}
\label{sec:2}

\begin{figure}[p]
\epsfsize=100pt \epsfbox[47 430 480 650]{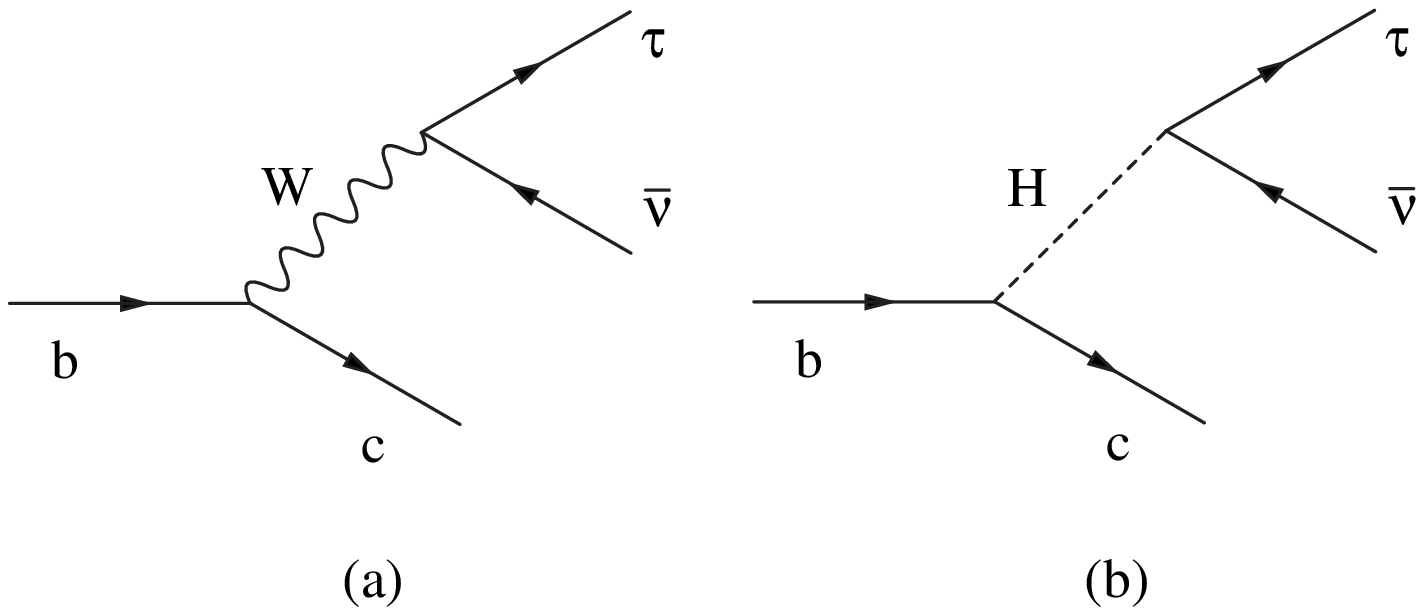}
\vspace{0pt}
\caption{Quark-level diagrams for the transition $b\to c \tau \overline{\nu}$
in a two-Higgs doublet model.}
\label{fig:b2c}
\end{figure}

The two diagrams which contribute to the decay
$B$$\to$$D\tau\overline{\nu}$ in a type II 2HDM are shown
in Fig.~\ref{fig:b2c}.  The amplitude corresponding to the SM 
$W$-exchange diagram (Fig.~\ref{fig:b2c}(a)) is given by
\equation
	{\cal M}_{\rm SM} = -2\sqrt{2}G_FV_{cb}
		\langle D(p^\prime)|\overline{c}_L\gamma^{\mu}b_L|B(p)\rangle 
		\overline{\tau}_L(p_{\tau})\gamma_\mu\nu_L(p_\nu),
\label{eq:smamp}
\endequation
where $\psi_{L}$$\equiv$$\frac{1}{2}(1-\gamma^5)\psi$.
The matrix element of the axial vector current in the above expression
is identically zero since one cannot form 
an axial vector using only $p$ and $p^\prime$.
The vector current matrix element may be expressed
in terms of two form factors, $F_0$ and $F_1$, which are defined
as follows:
\equation
	\langle D(p^\prime)|\overline{c}\gamma^{\mu}b|B(p)\rangle =
		F_1(t)\left[(p+p^\prime)^\mu-\frac{m_B^2-m_D^2}{q^2} q^\mu
		\right] + F_0(t)\frac{m_B^2-m_D^2}{q^2} q^\mu ,
\label{eq:formdef}
\endequation
with $q=p-p^\prime$ and $t=q^2/m_B^2$.  
The form factors $F_0$ and $F_1$ are
normalized such that $F_0(0)$$=$$F_1(0)$.  There is thus
no singularity at $q^2$$=$$0$.  

The parametrization of the form factors given in Eq.~(\ref{eq:formdef})
is particularly well-suited for our purposes since $F_0(t)$ and $F_1(t)$
may be associated with the spin-0 and spin-1 components of the
exchange particles,
respectively~\cite{wirbel}.  The contribution to the total amplitude
coming from the (spin-0) charged Higgs diagram
(Fig.~\ref{fig:b2c}(b)) may then be included by 
the following replacement in the SM expression for the amplitude:
\equation
	F_0(t)\rightarrow F_0(t)\left(1+\delta_H(t)\right).
\endequation
The function $\delta_H(t)$ is given by
\equation
	\delta_H(t) = -\left(\frac{\tan\beta}{m_H}\right)^2
		\frac{m_b m_B^2 t}{(m_B-m_D)}
		\left(1+\frac{m_c}{m_b}\cot^2\beta\right)
		\frac{F_S(t)}{F_0(t)},
\endequation
where the scalar form factor $F_S(t)$ is defined by
\equation
	\langle D(p^\prime)|\overline{c}b|B(p)\rangle =
		(m_B+m_D)F_S(t) .
\endequation

It is now straightforward to work out the expression for the differential
partial width in terms of these form factors.  Let us first define the 
following dimensionless quantities:
\equation
	r_D = \frac{m_D^2}{m_B^2}\; ,\;\;\;\;\;
	r_\tau = \frac{m_\tau^2}{m_B^2}\; .
\endequation
The expression for the width is then
\equation
     \frac{d\Gamma(B\to D\tau\overline{\nu})}{dt}
	   = \frac{G_F^2|V_{cb}|^2m_B^5}{128\pi^3} \rho(t) ,
\label{eq:width}
\endequation
where the dimensionless Dalitz density, $\rho(t)$, may be decomposed into 
spin-0 and spin-1 contributions as follows,
\equation
	\rho(t) = \left(1+\delta_H(t)\right)^2\rho_0(t) 
		+ \rho_1(t),
\endequation
with
\begin{eqnarray}
	\rho_0(t) & = & \frac{r_\tau}{t}
		\left[F_0(t)\right]^2 
		\left(1-\frac{r_\tau}{t}\right)^2(1-r_D)^2
		\lambda^{\frac{1}{2}}(1,r_D,t),\\
	\rho_1(t) & = & \frac{2}{3}
		\left[F_1(t)\right]^2
		\left(1-\frac{r_\tau}{t}\right)^2
		  \left(1+\frac{r_\tau}{2t}\right)
		\lambda^{\frac{3}{2}}(1,r_D,t),
\end{eqnarray}
and
\equation
	\lambda(a,b,c)=a^2+b^2+c^2-2(ab+ac+bc).
\endequation

The above expression for the differential width is a sum of two 
semi-positive definite terms, corresponding separately
to the spin-0 and spin-1 contributions.
The spin-0 contribution disappears in the limit 
$r_\tau$$\to$$0$, so that in this limit we recover
the familiar expression for the semileptonic decay to an electron
or muon.  This observation is actually very important, since the
distribution $d\Gamma_{(e,\mu)}/dt$ is expected to be measured
very precisely at the $B$ factories which are currently under construction.
Such a measurement would yield valuable information regarding the
distribution $d\Gamma_{(\tau)}/dt$, since 
\equation
	\frac{d\Gamma_{(\tau)}}{dt} =
		\frac{d\Gamma_{(e,\mu)}}{dt} 
		\left(1-\frac{r_\tau}{t}\right)^2
		\left[\left(1+\frac{r_\tau}{2t}\right)
		+\frac{3 r_\tau}{2t}\frac{(1-r_D)^2}{\lambda(1,r_D,t)}
		\zeta^2(t)\left(1+\delta_H(t)\right)^2\right],
	\label{eq:etaureln}
\endequation
where $r_\tau\leq t\leq\left(1-\sqrt{r_D}\right)^2$ and
\equation
	\zeta(t) \equiv \frac{F_0(t)}{F_1(t)} .
\endequation
Thus, the measurement of the differential distribution for the decays
$B\to D (e,\mu)\overline{\nu}$ may be used to {\em predict} 
the SM distribution for
the decay into the $\tau$, up to the function $\zeta(t)$.  As
we shall see in the next section, $\zeta(t)$ may be calculated
within the context of Heavy Quark Effective Theory with a relatively
small uncertainty.

In the remainder of this work we examine the $t$
distribution in Eq.~(\ref{eq:width}) in detail and evaluate 
its sensitivity to a Higgs signal.

\section{Theoretical uncertainties in the rate and in
the differential distribution}
\label{sec:3}

Our goal in this section is to estimate the theoretical
uncertainties in the distribution $d\Gamma_{(\tau)}/dt$
and in the integrated width
if the analogous distribution $d\Gamma_{(e,\mu)}/dt$ has
been measured very accurately.  The first step is to
evaluate the form factors $F_0$, $F_1$ and $F_S$ in
the heavy quark symmetry limit using the results
of Heavy Quark Effective Theory (HQET).
Since in this limit all of the form factors have a known
dependence on a single universal function -- the Isgur-Wise
function~\cite{isgurwise} -- one is potentially in very good shape for 
trying to disentangle the Higgs contribution from the SM 
contribution in the $t$ distribution.
The symmetry-breaking corrections to this picture introduce
theoretical uncertainties into the calculation of the function
$\zeta(t) = F_0(t)/F_1(t)$.  It is this function which
determines the shape of the distribution $d\Gamma_{(\tau)}/dt$
and whose uncertainties we shall need to estimate.

It is natural in the context of HQET to define quantities in terms
of the meson velocities instead of in terms of their momenta.  The 
matrix element of the hadronic vector current is then usually written
as
\equation
	\langle D(v^\prime)|\overline{c}\gamma^{\mu}b|B(v)\rangle =
		\sqrt{m_Bm_D}\left[h_+(w)(v+v^\prime)^\mu+
			h_-(w)(v-v^\prime)^\mu \right] ,
\label{eq:formdefhqet}
\endequation
where $v^\mu$$=$$p^\mu/m_B$, $v^{\prime\mu}$$=$$p^{\prime\mu}/m_D$ and
\equation
	w \equiv v\cdot v^\prime = \frac{1+r_D-t}{2\sqrt{r_D}} .
\endequation
Comparing the expressions in Eqs.~(\ref{eq:formdefhqet}) and 
(\ref{eq:formdef}), we find the (exact) correspondence
\begin{eqnarray}
	F_0(t) & = & -\frac{1}{2\,r_D^{1/4}}\left[
		\left(\frac{t - (1+\sqrt{r_D})^2}{1+\sqrt{r_D}}
			\right)h_+(w) -
		\left(\frac{t - (1-\sqrt{r_D})^2}{1-\sqrt{r_D}}
			\right)h_-(w)\right] ,\\
	F_1(t) & = & \frac{1}{2\,r_D^{1/4}}\left[
		(1+\sqrt{r_D})h_+(w) - (1-\sqrt{r_D})h_-(w)\right] .
\end{eqnarray}

The formalism of HQET gives a self-consistent way to express
hadronic form factors in an expansion in powers of 
$\Lambda/m_Q$, where $m_Q$ represents the masses of the heavy quarks
involved in the transition and where $\Lambda$ represents a dimensionful
quantity which is generically of order $\Lambda_{\rm QCD}$.  
For the meson form factors the first-order corrections are 
proportional to $\overline{\Lambda}/m_Q$, where $\overline{\Lambda}$
is defined as the difference between the meson and quark 
masses\footnote{Note that the pseudoscalar and 
vector mesons corresponding 
to a given heavy quark are degenerate
in mass at zeroth order in the heavy quark expansion.
$M_M$ is thus not the mass of any particular {\em physical} 
meson~\cite{neubert1}.}:
\equation
	\overline{\Lambda} = M_M - m_Q .
	\label{eq:lambardef}
\endequation
To leading order in the $1/m_Q$ expansion all of the form
factors may be expressed in terms of the universal Isgur-Wise function,
$\xi(w)$, which satisfies the normalization condition
$\xi(1)$$=$$1$.  In the heavy quark symmetry limit, and ignoring short-distance
QCD corrections, one finds that $h_+(w)$$\to$$\xi(w)$ and
$h_-(w)$$\to$$0$, so that
\begin{eqnarray}
	F_0(t) & \stackrel{HQS}{\longrightarrow} & 
		-\frac{1}{2\,r_D^{1/4}}
		\left(\frac{t - (1+\sqrt{r_D})^2}{1+\sqrt{r_D}}
			\right)\xi(w) ,
	\label{eq:f0hqs} \\
	F_1(t) & \stackrel{HQS}{\longrightarrow} & 
		\frac{1+\sqrt{r_D}}{2\,r_D^{1/4}}\xi(w) .
	\label{eq:f1hqs} 
\end{eqnarray}
Similar considerations for the scalar matrix element yield
\equation
	F_S(t) \stackrel{HQS}{\longrightarrow} F_0(t) .
	\label{eq:fsf0}
\endequation
The above expressions receive corrections due to the finite masses
of the heavy quarks.  The $1/m_Q$ corrections to $h_\pm$
have been considered in detail in Ref.~\cite{neubert1}, while the 
corrections to the scalar matrix element do not appear to have been
calculated.  For this reason we will, for the purpose of estimating
the theoretical errors, take the relation in Eq.~(\ref{eq:fsf0}) to
be exact.  This will not lead to significant errors in attempting
to bound small values of $\tan\beta/m_H$.  Under this
assumption, $\delta_H(t)$ takes the simple form
\equation
	\delta_H(t) \simeq -\left(\frac{\tan\beta}{m_H}\right)^2
		\frac{m_b m_B^2 t}{(m_B-m_D)} ,
\endequation
where we have also dropped the term proportional to $m_c$,
since its contribution to the amplitude is typically very small
for the range of $\tan\beta$ which we will be considering.

The form factors $h_\pm$ receive both short- and long-distance corrections.
The short-distance corrections are embodied in the Wilson
coefficients and may be calculated reliably using 
perturbation theory and renormalization group evolution.  They typically 
give corrections to the tree-level results which are on the order of 
$10\%$~\cite{neubert1}.  The long-distance corrections
are intrinsically non-perturbative and give rise to new
sub-leading universal functions.  At order $1/m_Q$ there are four such
functions~\cite{lukethm}.  The corrections to the results obtained in the
heavy quark symmetry limit may be taken into account by writing
\equation
	h_\pm(w) \equiv N_\pm(w) \xi_{\rm ren}(w) ,
\endequation
where the renormalized Isgur-Wise function is defined such that it
still satisfies $\xi_{\rm ren}(1)$$=1$.  The explicit expressions
for $N_\pm$ to order $1/m_Q$ may be found in Ref.~\cite{neubert1}.
Their values at zero recoil ($w$$=$$1$) are of particular interest
since one may use this information to extract $V_{cb}$ from 
the differential distribution for $B\to D (e,\mu)\overline{\nu}$.
It is known that $N_+$ is protected 
by Luke's theorem~\cite{lukethm} and thus does not receive any
$1/m_Q$ corrections at zero recoil.  $N_-$ does receive $1/m_Q$ 
corrections at zero recoil, but the resulting contributions to
the decay rate are parametrically suppressed by the 
factor $[(m_B-m_D)/(m_B+m_D)]^2\sim 0.23$~\cite{volshif,neubsup,lnn}.
The uncertainties in the decay rate at zero recoil due to $N_-$ 
are thus on the order of a few percent, which is about the same size 
as the expected $1/m_Q^2$ corrections.

We are now in a position to calculate the function $\zeta(t)$, the ratio
of $F_0$ and $F_1$, which appears in the expression for the differential
distribution given in Eq.~(\ref{eq:etaureln}).  Since
$|N_-|$$\ll$$|N_+|$, as will be clear {\em a posteriori},
it is an excellent approximation to expand $\zeta(t)$
in powers of $N_-/N_+$.  To first order this gives
\equation
	\zeta(t) \simeq \zeta_\infty(t) + \frac{4t\sqrt{r_D}}
		{(1-r_D)(1+\sqrt{r_D})^2}\frac{N_-(w)}{N_+(w)} ,
	\label{eq:zetaapprox}
\endequation
where 
\equation
	\zeta_\infty(t) = \frac{(1+\sqrt{r_D})^2-t}{(1+\sqrt{r_D})^2}
\endequation
is the result obtained in the limit of heavy quark symmetry.

The theoretical error associated with $\zeta(t)$ is actually remarkably
small.  The main reason for this is that there is an 
approximate accidental cancellation
in the expression for $N_-$ which tends to make it a very small number, to
leading order in $1/m_Q$~\cite{neubert1}.  
The resulting uncertainties due to our ignorance
of the forms of the sub-leading universal functions tend then
also to be small (on the order of several percent).  By way of
contrast, the uncertainties in $N_+$ are rather large 
(on the order of $20\%$), but these have a
negligible effect on the ratio $N_-/N_+$.  
This means that while the current predictions for $F_0(t)$ and 
$F_1(t)$ have large theoretical uncertainties, the prediction
for the {\em ratio} of $F_0(t)$ and $F_1(t)$ has a relatively small
theoretical uncertainty.  Thus, once $F_1(t)$ has been determined 
experimentally, $F_0(t)$ is also known quite well.  The 
ratio $N_-/N_+$ may be estimated by using the 
forms predicted for the sub-leading functions in specific model
calculations.  These functions have been calculated using QCD sum rules
in Refs.~\cite{neubsup,lnn,neubert1}.  We have studied $N_-/N_+$ 
numerically by allowing the sub-leading functions to vary over the regions
suggested in the plots\footnote{We have allowed the sub-leading function
$\chi_1^{\rm ren}(w)$ to vary over the larger shaded region shown in
Fig.~5.5 in Ref.~\cite{neubert1}.} in Ref.~\cite{neubert1} and by allowing
$\overline{\Lambda}$ to vary in the range
\equation
	0.4\; {\rm GeV} \leq \overline{\Lambda} \leq 0.6\; {\rm GeV} .
\endequation
We have, for consistency, taken the heavy quark masses to be 
defined in terms of $\overline{\Lambda}$ through Eq.~(\ref{eq:lambardef}), 
setting $m_b=4.8$ GeV and $m_c=1.45$ GeV when $\overline{\Lambda}=0.5$ GeV.
The resulting range for $N_-/N_+$ is then given by 
$-0.06 \leq N_-(w)/N_+(w) \leq 0.0$, 
for $1.0\leq w\leq 1.6$.  In order to account for the unknown $1/m_Q^2$
corrections, we conservatively double the size of this region.  We thus
estimate the range of theoretical uncertainty in $N_-/N_+$ to be
\equation
	-0.09 \leq \frac{N_-(w)}{N_+(w)} \leq 0.03 .
\endequation

It is now straightforward to use Eq.~(\ref{eq:etaureln}) to 
determine the differential width
for $B \to D \tau \overline{\nu}$ once that for 
$B \to D (e,\mu) \overline{\nu}$ is known.  The resulting curve
will have a theoretical uncertainty determined by the uncertainty 
in the ratio $N_-/N_+$.  We illustrate this
procedure in Fig. 2 by plotting $(1+\delta_H(t))^2\rho_0(t)$ (shaded bands)
and $\rho_1(t)$ (solid line)
for the SM and for 2HDM's with
$\tan\beta/m_H = 0.06$, $0.25$, and $0.35$ GeV$^{-1}$.  The spin-1
contribution to the width, $\rho_1(t)$, is independent of $\tan\beta/m_H$
and is assumed to have been determined experimentally from the decays to
the lighter leptons.  The spin-0 contribution,  $(1+\delta_H(t))^2\rho_0(t)$,
is determined from the
spin-1 contribution by using the relation $F_0(t) = \zeta (t) F_1(t)$,
with $\zeta(t)$ as given in Eq.~(\ref{eq:zetaapprox}).
For the purposes of this plot we have used the simple 
heavy quark symmetry relation for
$F_1(t)$ given in Eq.~(\ref{eq:f1hqs}), taking\footnote{This 
form for $\xi(w)$ is consistent with the current
experimental situation~\cite{expxi}.} $\xi(w) = 1 - 0.75\times(w-1)$.
Note that the spin-0 contribution is qualitatively quite 
different for $\tan\beta/m_H = 0.25$ GeV$^{-1}$ and 
$\tan\beta/m_H = 0.35$ GeV$^{-1}$.
These values fall on either side of the ``critical value'',
$\tan\beta/m_H \sim 0.3$ GeV$^{-1}$, 
for which the integrated width is at a minimum.
 From this plot we estimate that the current theoretical uncertainty
would allow one to use the differential distribution in 
$B\to D \tau \overline{\nu}$ to rule out a 2HDM with $\tan\beta/m_H > 0.06$
GeV$^{-1}$.

\begin{figure}[p]
\epsfsize=100pt \epsfbox[55 200 510 600]{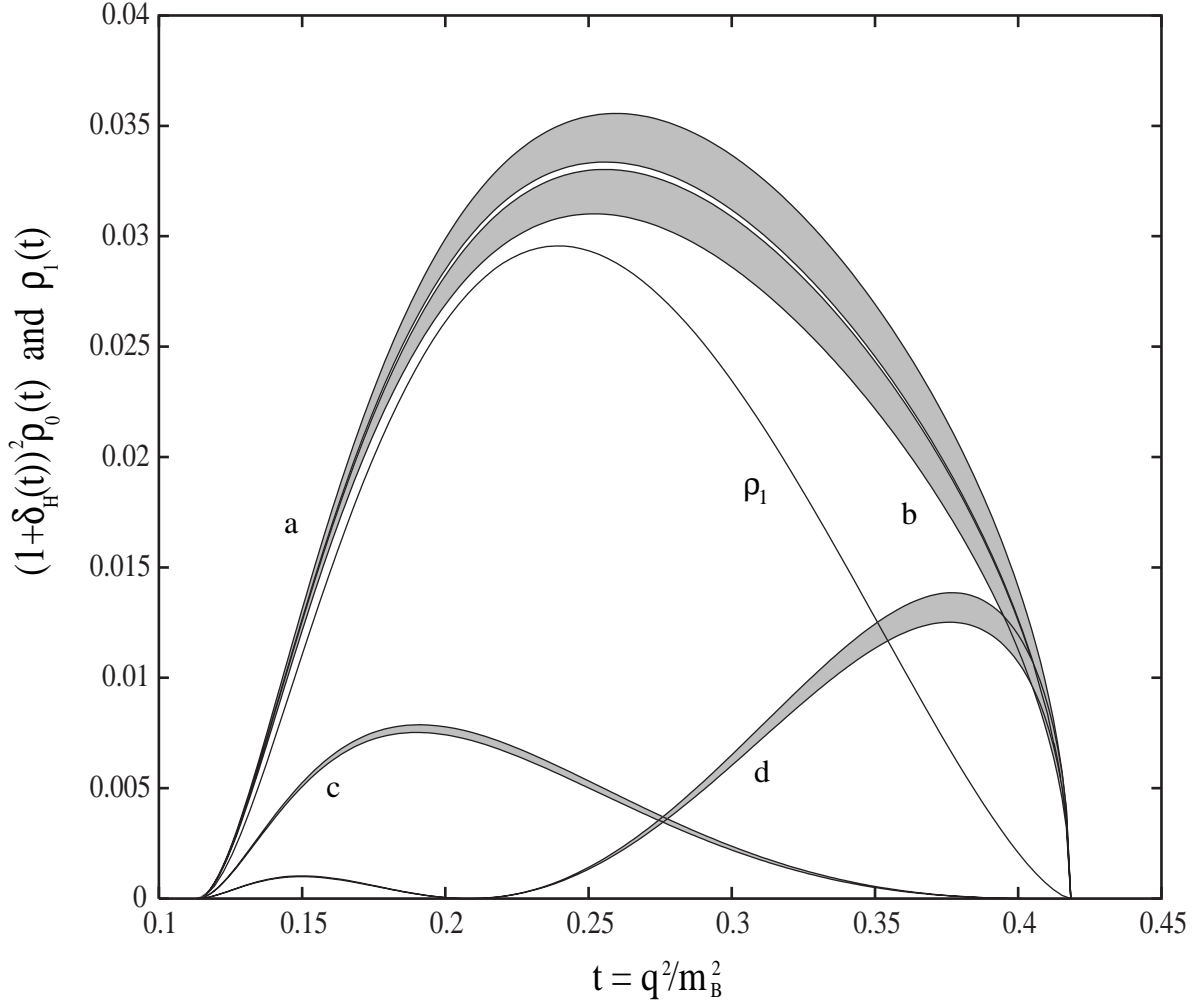}
\vspace{0pt}
\caption{Spin-0 and spin-1 contributions to the differential distribution 
in the standard model and in two Higgs doublet models
with various values of $\tan\beta/m_H$.  The shaded bands 
labeled a, b, c and d correspond
to the spin-0 contribution, $(1+\delta_H(t))^2\rho_0(t)$,
with $\tan\beta/m_H = 0$ (the SM),
$0.06$ GeV$^{-1}$, $0.25$ GeV$^{-1}$ and $0.35$ GeV$^{-1}$, respectively.
The solid curve corresponds to the spin-1 contribution, $\rho_1(t)$.
The shaded regions indicate the current theoretical uncertainty 
in $(1+\delta_H(t))^2\rho_0(t)$ (due to the uncertainty in
$\zeta(t)=F_0(t)/F_1(t)$) if $\rho_1(t)$ is known exactly.  
Uncertainties in $F_S(t)$ have not been included in this plot.}
\label{fig:rho}
\end{figure}

\begin{figure}[p]
\epsfsize=100pt \epsfbox[65 190 510 600]{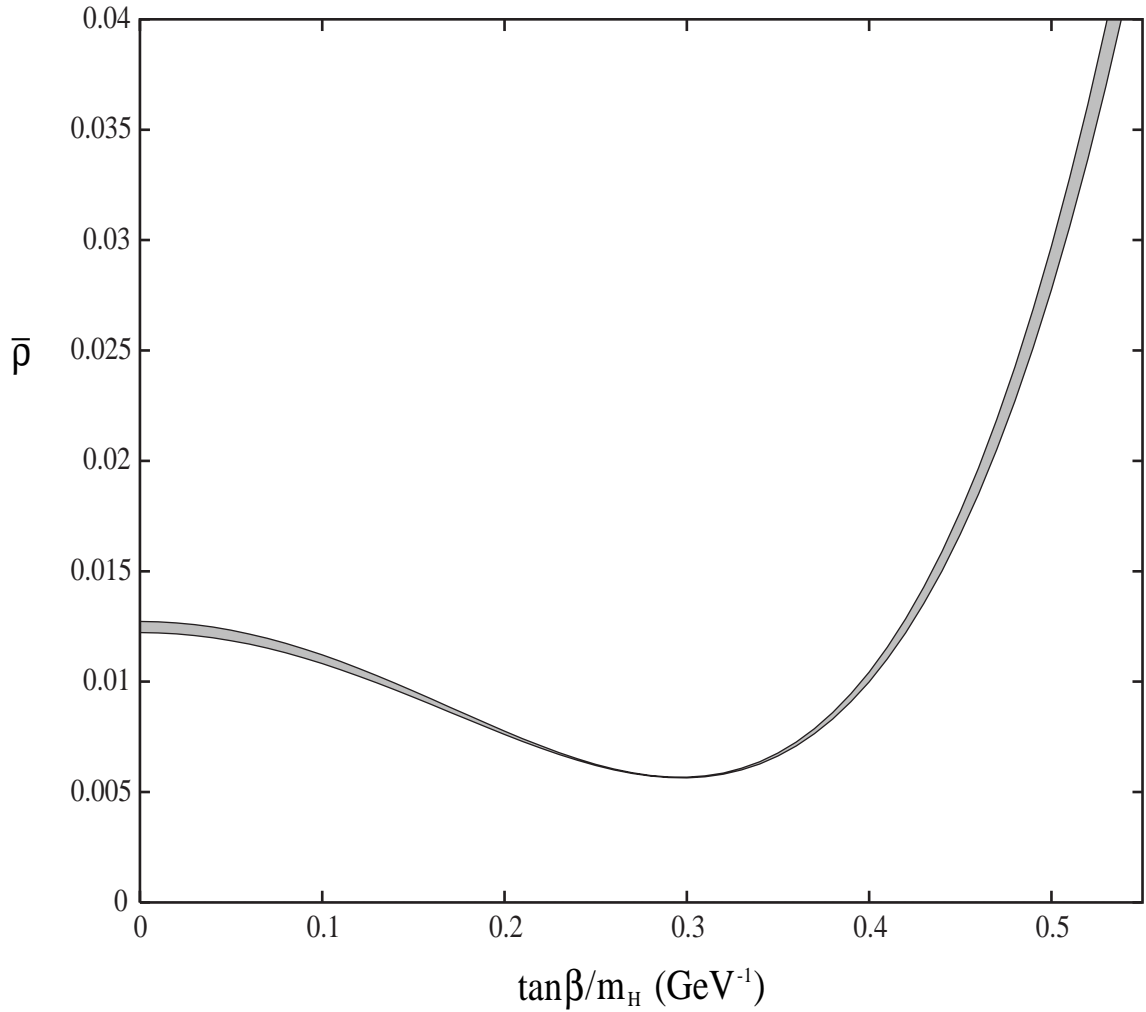}
\vspace{0pt}
\caption{Plot of the integrated width (normalized to 
$G_F^2|V_{cb}|^2m_B^5/128\pi^3$) as a function of $\tan\beta/m_H$.
As in Fig.~\ref{fig:rho}, the shaded region corresponds to the 
theoretical uncertainty in $\zeta(t)=F_0(t)/F_1(t)$ and does not take into
account the uncertainty in the scalar form factor, $F_S(t)$.}
\label{fig:rhoav}
\end{figure}

It is also useful to examine the behaviour of the integrated
width as a function of $\tan\beta/m_H$.  This behaviour is illustrated
in Fig.~\ref{fig:rhoav} in terms of the dimensionless quantity
$\overline{\rho}\equiv \int \rho(t)dt$, which is normalized
by the factor $G_F^2|V_{cb}|^2m_B^5/128\pi^3$ (see Eq.~(\ref{eq:width})).
As in Fig.~\ref{fig:rho}, the shaded band corresponds to the 
theoretical uncertainty in the ratio $\zeta(t)=F_0(t)/F_1(t)$ and
does not include the uncertainty in the form factor $F_S(t)$.
For the sake of illustration we have again used the simple
form for $F_1(t)$ given in 
Eq.~(\ref{eq:f1hqs}), taking $\xi(w) = 1 - 0.75\times(w-1)$.
The origin in this plot (i.e., the point $\tan\beta/m_H=0$)
corresponds to the SM.  For $\tan\beta/m_H>0$, the Higgs contribution
begins to intefere with the SM contribution, leading at first to 
a reduction in the width and then, for large values of 
$\tan\beta/m_H$, to an enhancement.  For $\tan\beta/m_H \gsim 0.45$ 
GeV$^{-1}$, the Higgs contribution completely dominates the width.
It is from this region that the current inclusive semi-tauonic 
bound on the ratio comes.  One may, in fact, compare our plot
with the analogous plot in the inclusive
case (see, for example, Fig.~1 in Ref.~\cite{ghn}).  Such a comparison
shows that while the two plots are qualitatively 
similar, the curve in the
present case has a more pronounced dip near 
$\tan\beta/m_H\sim 0.3$ GeV$^{-1}$, dropping to about $50\%$ of the
SM value at this point.  In the inclusive case the branching ratio at
the minimum drops to about $80-90\%$ of the SM value.
This feature illustrates a trend which we have already mentioned
above:  since the $B\to D^*\tau\overline{\nu}$ channel is not very
sensitive to the Higgs and since it has a relatively large branching ratio,
it tends to ``dilute'' the inclusive mode, making it less sensitive 
to the Higgs contribution.
Fig.~\ref{fig:rhoav} also illustrates why it is useful to have additional
information besides the integrated width.  If, for example,
the measured width is near or below the SM value, there will
alway be a two-fold degeneracy in the corresponding
value of $\tan\beta/m_H$.  The differential distribution may be
used to differentiate between the two values, however, since 
this distribution
is qualitatively different for values of $\tan\beta/m_H$
above and below the critical value.  The $q^2$ distribution
could also be extremely useful in ruling out small values
of $\tan\beta/m_H$, since the integrated width is quite flat
in this region.

The shaded bands shown in Figs.~\ref{fig:rho} and \ref{fig:rhoav}
correspond to our estimates of the current theoretical
uncertainties in $F_0(t)$ and $F_1(t)$ and do not take into account
the uncertainties associated
with $F_S(t)$.  These uncertainties could in 
principle be on the order of $10-20\%$,
although they could also be small, as was the case for
$N_-$.  It is beyond the scope of this paper to provide a more quantitative
estimate for the uncertainty in $F_S(t)$.  Let us note, however, that
these uncertainties have a negligible effect for small values of 
$\tan\beta/m_H$ and thus do not affect one's ability to rule out
a 2HDM with a small value of $\tan\beta/m_H$.  
Should one observe evidence for a non-zero value
of the ratio $\tan\beta/m_H$, one would clearly want to calculate $F_S(t)$
more carefully in order to precisely determine this ratio.

\section{Using the optimized weighting procedure}
\label{sec:4}

We have so far considered the theoretical uncertainties which arise
in the calculation of the differential distribution for 
$B\to D\tau\overline{\nu}$.  These uncertainties determine --
in the limit of infinite experimental statistics -- our ability to 
distinguish the standard model from a two Higgs doublet model
with a given value of $\tan\beta/m_H$.  Let us now turn the situation
around and ask the following question:  Suppose all of the form
factors could be determined precisely by some means\footnote{
This is not an unreasonable assumption, since the form factors can in 
principle be determined on the lattice~\cite{lattice}.  Even without
lattice results, the sub-leading universal functions may eventually be
constrained by precision measurements in some of the other $B$ decay
channels.}.  How well could
one then differentiate between the SM and a 2HDM with a finite
number of events?
This question may be answered by using the optimized weighting procedure
which has recently been discussed in Refs.~\cite{atsoni,ggh}.
This procedure provides the most efficient way (with regard to
statistical uncertainties) to analyze the experimental
data in order to differentiate between the two models\footnote{
Note that, while this technique provides the best
way to minimize statistical uncertainties, in an actual experimental
setting one will have to consider the effect of systematic errors as well.}.

In this section we will briefly review the optimal observables method
and then apply it to the problem at hand, which is to distinguish
between the SM and a 2HDM.  Let us first review the method.
Suppose one has a distribution of the form 
\equation
	{\cal O}(\phi) = \sum_i c_i f_i(\phi) ,
\endequation
where $\phi$ represents some collection of kinematical variables,
the functions $f_i(\phi)$ are known functions of those variables
and the constants $c_i$ parametrize the different models which one
wishes to investigate.  It may be demonstrated that the optimal 
observables method provides the most efficient way in which to 
extract the coefficients $c_i$.
This technique was first discussed in Ref.~\cite{atsoni} for the 
case in which $c_1=1$ and $c_2=\lambda$, with $\lambda$ being some
small number.  We will use the generalized version presented in 
Ref.~\cite{ggh}, since for large values of $\tan\beta/m_H$ the
$c_i$ need not be small.

The main goal of the optimal observables approach is to find
the optimal set of functions $w_i(\phi)$ such that
\equation
	c_i = \int w_i(\phi){\cal O}(\phi) d\phi .
\endequation
As shown in Ref.~\cite{ggh}, this set is given by
\equation
	w_i(\phi) = \frac{\sum_j X_{ij}f_j(\phi)}{{\cal O}(\phi)},
\endequation
where
\begin{eqnarray}
	X_{ij} & = & M^{-1}_{ij} , \\
	M_{ij} & \equiv & \int \frac{f_i(\phi)f_j(\phi)}{{\cal O}(\phi)}
		d\phi .
\end{eqnarray}
The coefficients $c_i$ are then given by
\equation
	c_i = \sum_j M_{ij}^{-1} \left(\int f_j(\phi) d\phi\right) .
\endequation
The experimental task reduces to measuring the elements of the matrix
$M$.  The statistical error associated with this procedure is embodied
in the $\chi^2$ function, defined by
\equation
	\chi^2 = \frac{N}{\sigma_T} \sum_{i,j} (c_i-c_i^0)(c_j-c_j^0) M_{ij},
	\label{eq:chi2}
\endequation
where the $c_i^0$ represent the measured values of the coefficients,
$N$ is the total number of events and $\sigma_T = \int {\cal O}(\phi) d\phi$.

The above procedure is straightforward to implement in our case.
Dropping the dimensionful prefactor in Eq.~(\ref{eq:width}),
we write the differential distribution in terms of the Dalitz
density as follows,
\equation
	\rho(t) = \sum_i c_i f_i(t),
\endequation
where
\begin{eqnarray}
	f_1(t) & = & \rho_0(t) + \rho_1(t) ,\\
	f_2(t) & = & -\frac{2m_bt}{(m_B-m_D)} \rho_0(t) ,\\
	f_3(t) & = & \frac{m_b^2t^2}{(m_B-m_D)^2} \rho_0(t) ,
\end{eqnarray}
and
\equation
	c_1 = 1, \;\;\;\;\;\; c_2 = \alpha_H , \;\;\;\;\;\; c_3 = \alpha_H^2 .
	\label{eq:ci}
\endequation
The dimensionless parameter $\alpha_H$ is defined as
\equation
	\alpha_H = \left(\frac{m_B\tan\beta}{m_H}\right)^2 .
\endequation
We may now use the machinery of the optimized weighting procedure
in order to calculate the statistical errors for a given model
(i.e., for a given value of $\tan\beta/m_H$)
and a given number of experimental events.  We take as input 
some value of $\tan\beta/m_H$, calculate the elements of the matrix
$M$ and then perform the sum in Eq.~(\ref{eq:chi2}).
The $c_i^0$ in this expression are given by the input
values themselves and the $c_i$ are as indicated\footnote{Note that 
we set $c_1=c_1^0=1$, since we assume that the SM $W$-exchange 
contribution will always be present.} in Eq.~(\ref{eq:ci}).
$\chi^2$ is thus a polynomial in even powers of $\tan\beta/m_H$
and is zero at the input value.  The $n\sigma$ error
for a given experiment is simply gotten by setting 
$\chi^2$$=$$n^2$~\cite{ggh}.

\begin{figure}[p]
\epsfsize=100pt \epsfbox[55 210 505 610]{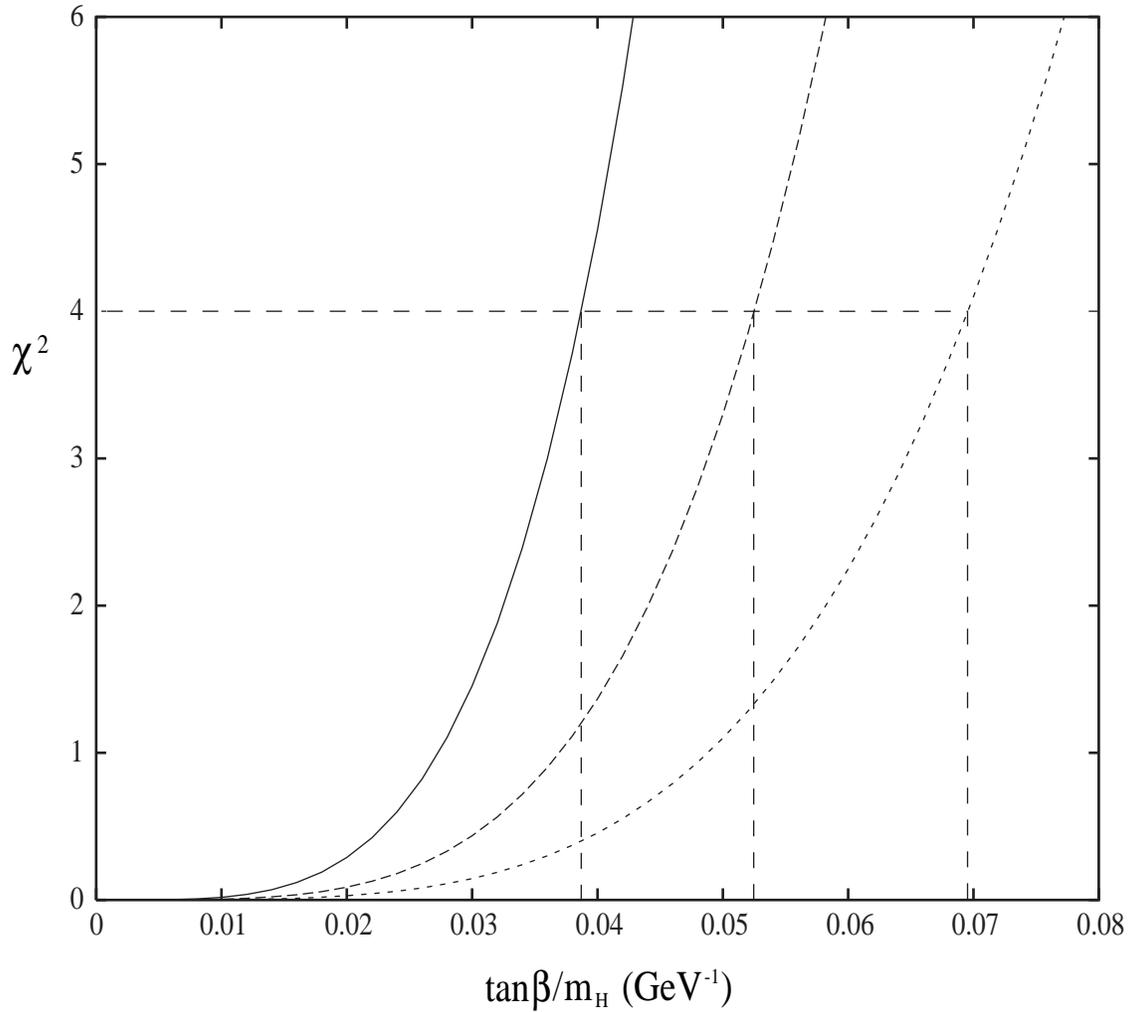}
\vspace{0pt}
\caption{$\chi^2$ for the optimized weighting technique
as a function of $\tan\beta/m_H$.  The ``input model'' in this case
is the standard model (i.e., $\tan\beta/m_H$$=$$0$).  The solid,
dashed and short-dashed curves correspond to $10^4$, $3\times 10^3$ and
$10^3$ events, respectively.  The vertical dashed lines indicate
the $2\sigma$ upper bound on $\tan\beta/m_H$ in each case, with the
regions to the right of the dashed lines being excluded.}
\label{fig:chi1}
\end{figure}

For our numerical analysis we have used the 
simple heavy quark symmetry forms for $F_0$,
$F_1$ and $F_S$ (see 
Eqs.~(\ref{eq:f0hqs})--(\ref{eq:fsf0})), again taking
$\xi(w) = 1-0.75\times(w-1)$.
In Fig.~\ref{fig:chi1} we plot $\chi^2$ as a function of $\tan\beta/m_H$
for the case in which the input model is the SM ($\tan\beta/m_H = 0$).  
The solid, dashed and short-dashed 
curves correspond to $N=10^4$, $3\times 10^3$ and $10^3$ 
events, respectively.  The dashed vertical lines indicate the 
$2\sigma$ upper bounds which could be placed on $\tan\beta/m_H$ in
each case.  An approximate formula for this upper bound may
be obtained from the expression for $\chi^2$
by setting $\chi^2$$=$$4$ and truncating the polynomial
in $\tan\beta/m_H$ at the leading term.
This leads to the following approximate expression for the
$2\sigma$ upper bound:
\equation
	\tan\beta/m_H \lsim \frac{0.39}{N^{1/4}}\;\; {\rm GeV}^{-1} .
\endequation
For the three cases shown in the plot, this corresponds to
\equation
	\tan\beta/m_H \lsim \left\{\begin{array}{ll}
				0.039 \;\;{\rm GeV}^{-1},\;\;\;\;\; & N=10^4 \\
				0.052 \;\;{\rm GeV}^{-1}, & N=3\times 10^3 \\
				0.069\;\;{\rm GeV}^{-1}, & N=10^3 ,
				\end{array}\right.
	\label{eq:tanbmh3}
\endequation
in reasonable agreement with the exact results indicated in the plot.
The upper bounds determined in this way
would represent significant improvements
on the current limits, which are $0.46$ GeV$^{-1}$ and $0.38$ GeV$^{-1}$
(coming from the inclusive semi-tauonic and tauonic $B$ decays, respectively).
Recall, furthermore, that the current theoretical uncertainties 
in the form factors would limit the reach of even an ideal 
experiment (with infinite statistics) 
to about $0.06$ GeV$^{-1}$ (see Fig.~\ref{fig:rho}).  
We see from Eq.~(\ref{eq:tanbmh3}) that already with 
about $3\times 10^3$ events
one will have reached this limit.

\begin{figure}[p]
\epsfsize=100pt \epsfbox[60 200 505 600]{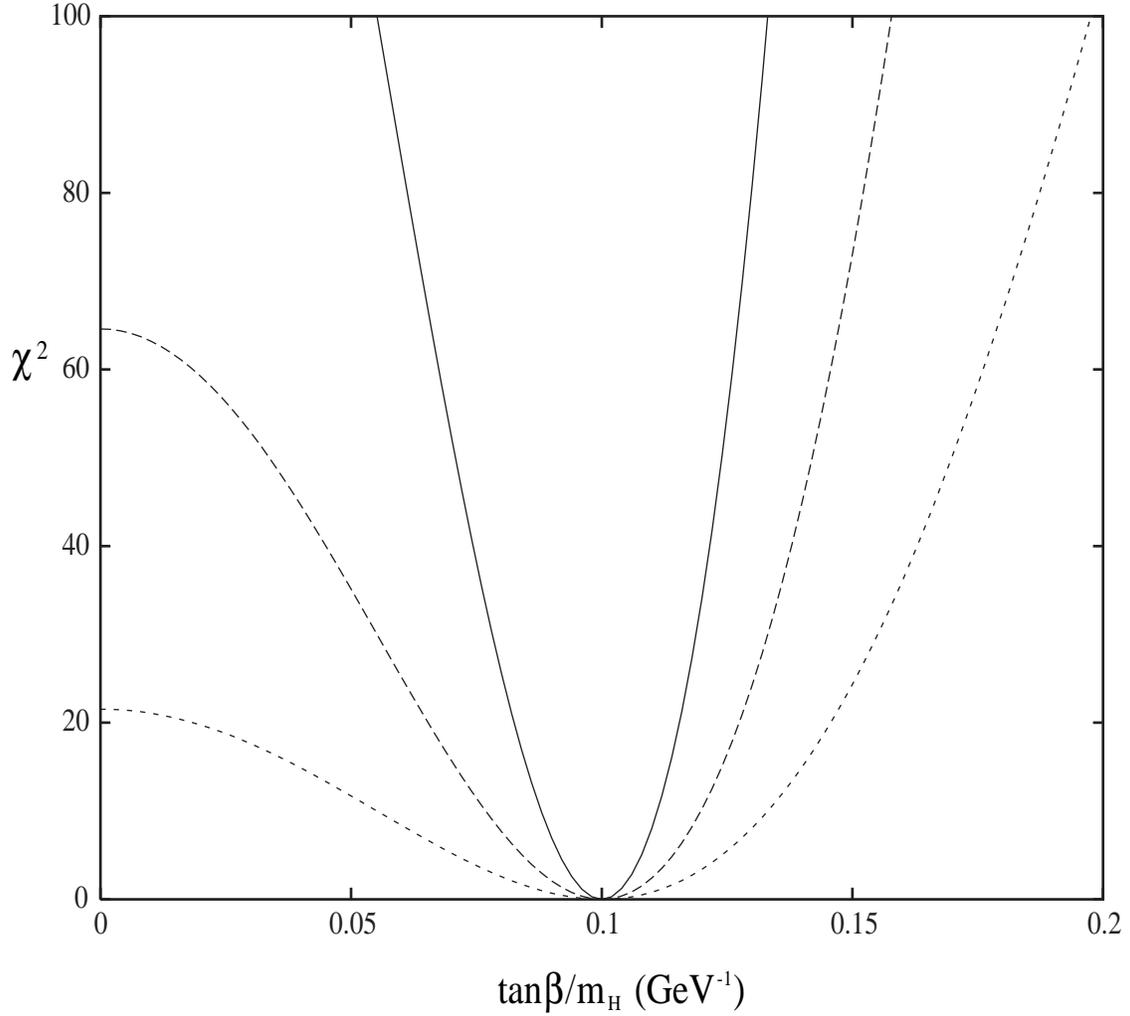}
\vspace{0pt}
\caption{$\chi^2$ for the optimized weighting technique
as a function of $\tan\beta/m_H$ in a two Higgs doublet model.
The ``input model'' in this case has
$\tan\beta/m_H=0.1$ GeV$^{-1}$.  The solid,
dashed and short-dashed curves correspond to $10^4$, $3\times 10^3$ and
$10^3$ events, respectively.  The intercept at $\tan\beta/m_H=0$
shows that for $10^3$ events one can distinguish this scenario
 from the standard model at approximately the $4.6\sigma$ level.}
\label{fig:chi2}
\end{figure}

This approach is also extremely sensitive to non-zero input values
of $\tan\beta/m_H$.  As an example, we have calculated $\chi^2$
for a 2HDM with $\tan\beta/m_H = 0.1$ GeV$^{-1}$.  The resulting
plot, given as a function of $\tan\beta/m_H$, is shown in 
Fig.~\ref{fig:chi2}.  The solid, dashed and short-dashed 
curves correspond again to $N=10^4$, $3\times 10^3$ and $10^3$ 
events, respectively.  The intercepts of the three
curves at $\tan\beta/m_H = 0$ determine how well one can differentiate
between this 2HDM and the SM in each case.  
Thus, for example, with about $10^3$ events one can 
differentiate between this model 
and the SM at approximately the $4.6\sigma$ level.

\section{Concluding remarks}
\label{sec:5}

In this paper we have examined the sensitivity of the exclusive
decay $B\to D\tau\overline{\nu}$ to the tree-level charged Higgs
contribution which is generic to type II two Higgs doublet models.
The $q^2$ distribution in this decay is extremely sensitive to 
the ratio $\tan\beta/m_H$ and may be used to appreciably improve
the existing upper bounds on this quantity or, as the case may be,
to measure a non-zero value.  We have shown that while the existing
theoretical uncertainties on this distribution are rather large, they
will be reduced significantly once the analogous distribution for
$B\to D(e,\mu)\overline{\nu}$ is measured more precisely.  We estimate
that, barring any further theoretical reductions in the uncertainty,
this would eventually (i.e., assuming an infinite number of experimental
events) allow one to rule out a 2HDM with
$\tan\beta/m_H>0.06$ GeV$^{-1}$.  We have also applied the optimized
weighting technique to the $q^2$ distribution in order to calculate
the minimum statistical uncertainty which could be attained
for an experiment with a given number of events.  The results of this
analysis are very encouraging, showing, for example, that with just 
$10^3$ events one could rule out a 2HDM with 
$\tan\beta/m_H>0.07$ GeV$^{-1}$.  With the same number of events
one could differentiate between a 2HDM with $\tan\beta/m_H=0.1$ GeV$^{-1}$
and the SM at the $4.6\sigma$ level.

In the present work we have not made any use of the spin of the final
state $\tau$.  In principle, this extra observable could be used to 
improve the sensitivity of a given experiment.  In practice, one
has to allow the $\tau$ to decay and study the distribution of its
decay products.  We have, in fact, used the optimized
weighting procedure to study these distributions
in the hadronic decay channels $\tau\to\pi\nu$ and $\tau\to\rho\nu$.
The differential distributions in these cases may be written as functions of
$q^2$, $E_h$ and $\cos\theta_h$, where $E_h$ and $\theta_h$ are the
energy and angle (with respect to the momentum of the $D$) of the meson
$h$ in the $\tau$$-$$\overline{\nu}$ rest-frame.  
(Recall that we wish to avoid observables
which depend explicitly on the momentum of the $\tau$.)  
The upshot of this analysis is that for a fixed number of events,
the upper limit on $\tan\beta/m_H$ is improved by at most about $5\%$.
In general, however, one can also expect a reduction in the number of events
since one is now considering a very specific decay mode of the $\tau$.
Thus one is likely not to gain anything at all.
It is thus our opinion that the $q^2$ distribution
represents perhaps the best tool which may
be used in order to search for a charged Higgs signal 
in $B\to D\tau\overline{\nu}$.

\acknowledgments
We would like to thank T. Blum, S. Dawson, F. Paige, L. Reina,
M. Tytgat and G.-H. Wu for helpful discussions.
This research was supported in part by the U.S. Department of Energy
under contract number DE-AC02-76CH00016.  
K.K. is also supported in part by the Natural Sciences and
Engineering Research Council of Canada.

\addtolength{\baselineskip}{-.3\baselineskip}

\end{document}